\documentclass{nature}
\usepackage{graphicx}

\title{Time-reversal symmetry breaking hidden order in Sr$_2$(Ir,Rh)O$_4$}

\author{Jaehong Jeong$^1$, Yvan Sidis$^1$, Alex Louat$^2$, V\'eronique Brouet$^2$ \& Philippe Bourges$^1$}

\begin{document}

\maketitle
 
\begin{affiliations}
 \item Laboratoire L\'eon Brillouin, CEA-CNRS, Universit\'e Paris-Saclay, CEA Saclay, 91191 Gif-sur-Yvette, France
 \item Laboratoire de Physique des Solides, Universit\'e Paris-Sud, Universit\'e Paris-Saclay, 91405 Orsay, France

\end{affiliations}

\begin{abstract}
 
Layered 5$d$ transition iridium oxides, Sr$_2$(Ir,Rh)O$_4$, are described as unconventional Mott insulators with strong spin-orbit coupling.  The undoped compound, Sr$_2$IrO$_4$, is a nearly ideal two-dimensional pseudospin-$1/2$ Heisenberg antiferromagnet, similarly to the insulating parent compound of high-temperature superconducting copper oxides. Using polarized neutron diffraction, we here report a hidden magnetic order in pure and doped Sr$_2$(Ir,Rh)O$_4$, distinct from the usual antiferromagnetic pseudo-spin ordering. We find that time-reversal symmetry is broken while the lattice translation invariance is preserved in the hidden order phase. The onset temperature matches that of the odd-parity hidden order recently highlighted using optical second harmonic generation experiments. The novel magnetic order and broken symmetries can be explained by the loop-current model, previously predicted for the copper oxide superconductors.

\end{abstract}

In the $5d$ layered perovskite material, Sr$_2$IrO$_4$, spin-orbit coupling and strong electron correlations combine to give rise to  a spin-orbit coupled Mott insulator with a pseudo-spin $J=1/2$ antiferromagnetic (AFM) state~\cite{BJ2008,BJ2009}.  It exhibits close structural~\cite{Ye2013,Dhital2013}, electronic~\cite{Wang2011,Arita2012} and magnetic~\cite{Ye2013,Dhital2013} similarities with the $3d$ layered perovskite material, La$_2$CuO$_4$, which evolves from a spin $S=1/2$ antiferromagnetic Mott insulator to a high temperature superconductor upon doping.  Doped Sr$_2$IrO$_4$ has then become a quite promising material to discover new states of matter, including unconventional superconductivity.

\begin{figure*}
\includegraphics[width=\textwidth,clip]{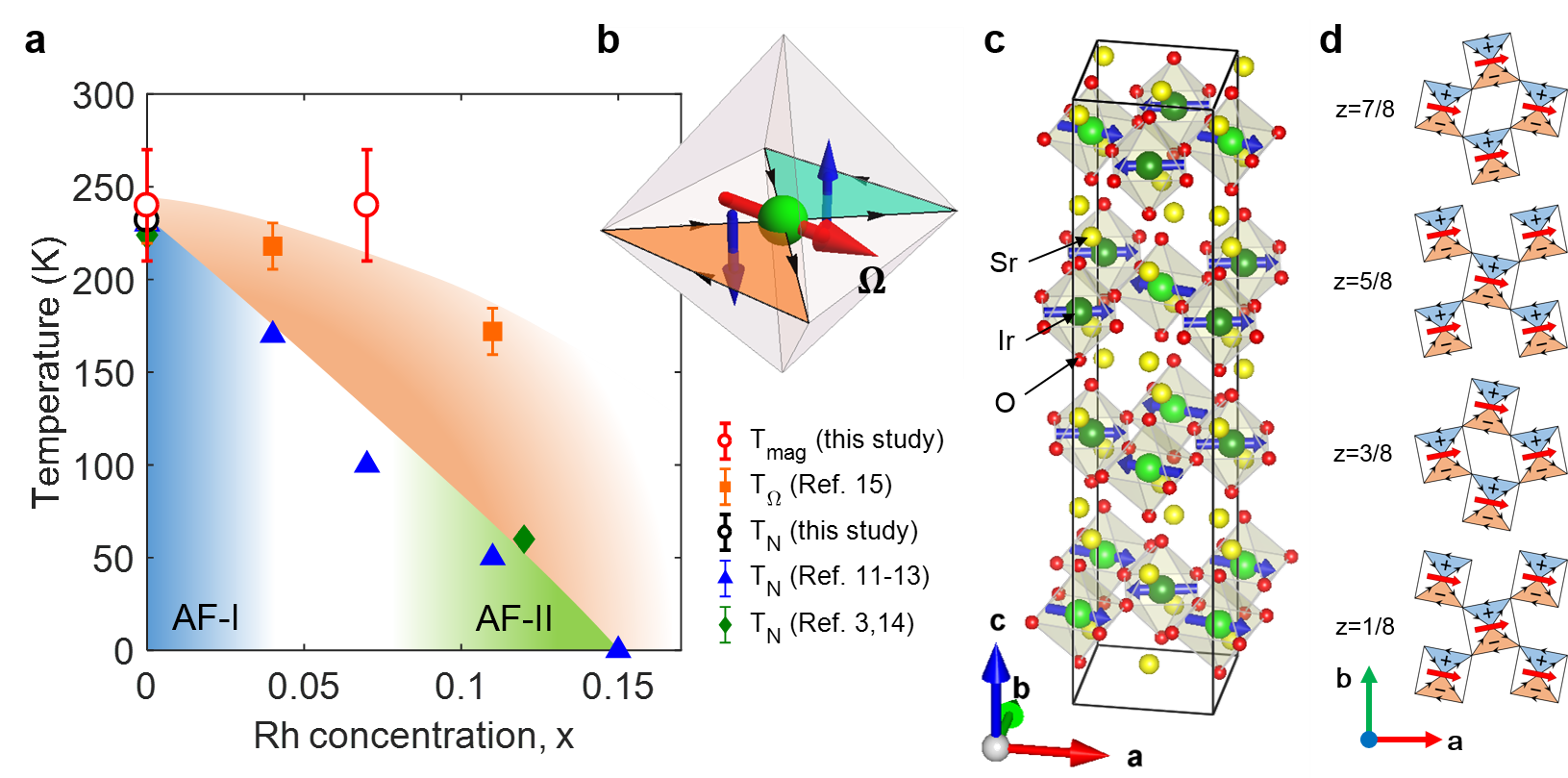}
\caption{\label{fig1}{\bf Magnetic phase diagram and loop-current order} 
(a) Magnetic phase diagram of Sr$_2$(Ir,Rh)O$_4$. The obtained transition temperatures, $T_\mathrm{N}$ and $T_\mathrm{mag}$ in this study are represented by black and red empty circles for the conventional AFM and the hidden magnetism preserving the lattice translation, respectively. The hidden broken symmetry phase, $T_\Omega$, observed by SHG~\cite{Zhao2016} (orange squares) is also represented as well as the AFM phase determined by magnetization measurements~\cite{Qi2012,Cao2016,Brouet} (blue triangles) and by neutron diffraction~\cite{Ye2013,Ye2015} (green diamonds). 
(b) A schematic picture for co-planar LC state in a single IrO$_6$ octahedron. The blue and red arrows denote magnetic moments generated by circulating currents and the anapole, $\Omega$, respectively.
(c) Atomic and AFM structures of Sr$_2$IrO$_4$ with a space group $I4_1/a$ (origin choice 2) (Ir atoms are represented in green, Sr in yellow and oxygen in red).
(d) The co-planar LC ordered state in the basal plane. The red arrows denote the anapole vector, $\Omega$, as shown in (b), and the plus/minus signs correspond to the orbital magnetic moments perpendicular to the $ab$-plane. This {\it nearly}-ferro-toroidal order preserves translational symmetry of lattice but breaks parity inversion, four-fold rotational and time-reversal symmetries.}
\end{figure*}

The crystal structure of the layered perovskite  Sr$_2$IrO$_4$ is characterized by the stacking of IrO$_2$ and SrO layers. The Ir$^{4+}$ ion is at the center of an oxygen octahedron, rotated by $\theta=11^\circ$ in the basal $ab$ plane. (Fig.~\ref{fig1}c). For a long time, Sr$_2$IrO$_4$ has been described as possessing  a tetragonal centrosymmetric structure with four-fold rotational symmetry about the $c$-axis, corresponding to a space group $I4_1/acd$~\cite{Huang1994}. However, a structural distortion exists even above room temperature as shown by optical second harmonic generation studies~\cite{Torchinsky2015} yielding to the space group $I4_1/a$ where the $c$- and $d$-glide planes are lost. That gives rise to additional weak Bragg spots such as $(1,0,2n+1)$ that have been reported using neutron diffraction~\cite{Ye2013,Dhital2013}. In this material, crystal field effects, spin-orbit coupling, Coulomb repulsion and the bending of the Ir-O-Ir bonds play an important role to understand electronic and magnetic properties.

Within the octahedral crystal field, the Ir $5d$ electronic levels split into e$_g$ and t$_{2g}$ states. Under strong spin-orbit coupling, the t$_{2g}$ states split into a $J=1/2$ doublet and $J=3/2$ quartet, so that, among the five $5d$ electrons of the Ir$^{4+}$ ion, only one remains in the $J=1/2$ doublet state. A large enough effective Coulomb repulsion finally localizes the $J=1/2$ electron and one is left with a $J=1/2$ pseudo-spin model. Below  $T_{\rm{N}}\sim$230 K~\cite{BJ2008}, an AFM order develops, characterized by a magnetic propagation wavevector $\mathbf{q}_m=(1,1,1)$ and   magnetic moments at the Ir sites aligned in the basal $ab$-plane ~\cite{Ye2013,Dhital2013,Boseggia2013} (Fig.~\ref{fig1}c). The directions of the staggered magnetic moments are tied to IrO$_6$ octaedra and follow their rotation, $\theta$, giving rise to a canting of the AFM structure in each IrO$_2$ layer with a ferromagnetic component along the $b$-axis. As shown in Fig.~\ref{fig1}c, the magnetic structure of Sr$_2$IrO$_4$ has a staggered stacking along the $c$-axis of the canted AFM layers (defined as $+$ or $-$ depending of the sign $\pm\theta$ of the tilt). In other words, the weak ferromagnetic component in each layer is stacked to give a $-++-$  structure along $c$-axis (AF-I phase)~\cite{Ye2013,Dhital2013}, so it cancels out globally.  However, the AFM interaction along the $c$-axis is rather weak, thus it is easily transformed to the $++++$ stacking order (AF-II phase)  by applying an external  magnetic field ~\cite{BJ2009}. At variance with the AF-I order, the AF-II stacking is then characterized by a magnetic propagation vector $\mathbf{q}_m=\mathbf{0}$~\cite{Clancy2014} and a net (but weak) ferromagnetic moment. Thorough that paper, we refer to these AFM phases, as the one depicted in Fig.~\ref{fig1}c, as conventional AFM phases. 

Interestingly, the AFM transition temperature is suppressed under the substitution of Rh for Ir. While the Rh substitution could be thought to be iso-electronic, it should be stressed that Rh is likely to play the role of an acceptor (Rh$^{3+}$/Ir$^{5+}$) and effectively give rise to a hole-doping of the material~\cite{Clancy2014}. Fig.~\ref{fig1}a shows the magnetic phase diagram in Sr$_2$Ir$_{1-x}$Rh$_x$O$_4$. From bulk magnetization~\cite{Qi2012,Cao2016,Brouet} and neutron diffraction measurements~\cite{Ye2013,Ye2015}, the conventional AFM transition decays  almost linearly as a function of Rh substitution up to $x_c\approx0.15$. Actually, x-ray studies~\cite{Clancy2014} indicate a slightly larger critical doping $x_c\approx0.17$ with the occurence of another transition at lower temperature when the magnetic correlation lengths are effectively diverging. Further, upon Rh-substitution, the AFM order undergoes a transition from the AF-I to the AF-II phase, even at zero magnetic field~\cite{Clancy2014,Ye2015}.

Recently, a hidden broken symmetry phase, developing prior to the AFM state, has been reported in Sr$_2$(Ir,Rh)O$_4$ using rotational anisotropy optical second harmonic generation (SHG) measurements~\cite{Zhao2016}. The hidden broken symmetry phase was observed distinctively at a few K above $T_\mathrm{N}$ for the pure sample, and far above for the doped systems~\cite{Zhao2016}.  
These data highlight an odd-parity hidden order as both  the inversion and four-fold rotational symmetries are broken below a temperature $T_\Omega$ distinct from the N\'eel temperature $T_\mathrm{N}$ (Fig.~\ref{fig1}a)~\cite{Zhao2016}. From the symmetry analysis, the SHG results could be in principle explained by a triclinic distortion of the crystal structure. However there is no experimental evidence using x-ray and neutron scattering of any structural distortion~\cite{BJ2009,Ye2013,Dhital2013}. It should be nevertheless stressed that these diffraction studies use too modest spatial resolution to definitively prove a lack of symmetry lowering. 

Alternatively, the SHG signal could be due to a magnetic ordering although there is no proof of time-reversal symmetry breaking at $T_\Omega$. Actually, the AF-I ground state of pure Sr$_2$IrO$_4$ preserves the parity inversion symmetry and thus cannot explain the SHG signal. Instead, a few magnetic point groups that preserve the translation symmetry of the lattice were proposed to account for the SHG signal, such as $2'/m$ or $m1'$~\cite{Zhao2016,DiMatteo2016}. In particular, a so-far non-observed AFM state of $2'/m$ symmetry, corresponding to a stacking $+-+-$ along the $c$ axis of AFM planes, would produce the SHG signal~\cite{DiMatteo2016}. 

Among the magnetic point groups, it is argued~\cite{Zhao2016} that the new broken symmetries can be caused by a loop current (LC) phase~\cite{Varma1997,Simon2002}, theoretically proposed to account for the pseudo-gap physics of superconducting cuprates. The existence of such a magneto-electric state has gained support from the detection in several cuprate families of its magnetic fingerprint by polarized neutron diffraction~\cite{Fauque2006,Li2008,Bourges2011,Sidis2013,Mangin2015}. Using the same technique for  Sr$_2$(Ir,Rh)O$_4$, we here report at the temperature of the odd-parity order,  $T_\Omega$, the appearance of a hidden magnetic order, which breaks time reversal symmetry while preserving lattice translation invariance. Among the magnetic models inferred from the SHG data~\cite{Zhao2016}, only the co-planar LC order~\cite{Varma1997,Simon2002} produces a magnetic diffraction pattern consistent with our polarized neutron data. Our results show that exotic magnetic orders with the same symmetry properties as the LC phase exist in both iridates and cuprates. 

{\bf Results}

Let us first describe the co-planar loop current order. It is characterized by two circulating currents turning clockwise and anti-clockwise within the same plane inside the IrO$_6$ octahedron (Fig.~\ref{fig1}b), and belongs to a $2'/m$ point group symmetry. It breaks time-reversal and inversion symmetries but not their product. The LCs produce two opposite orbital magnetic moments within each IrO$_6$ octahedron. A toroidal pseudo-vector, or anapole, is defined as an order parameter by $\mathbf{\Omega}=\sum_i\mathbf{r}_i\times\mathbf{m}_i$, where $\mathbf{r}_i$ and $\mathbf{m}_i$ stand for the position and the orbital magnetic moment, respectively, in one octahedron (Fig.~\ref{fig1}b).  It is similar to the toroidal moment in multiferroic systems~\cite{Spaldin2008}.  When the anapole vectors are parallel along the $c$-axis (ferro-toroidal coupling) as shown in Fig.~\ref{fig1}d, the ordered structure does not break translational symmetry but breaks parity inversion and four-fold rotational symmetries~\cite{Zhao2016}. Since the direction of the anapole is bound to the orientation of each IrO$_6$ octahedron, the resulting order is a {\it nearly}-ferro-toroidal order ({\it i.e.} weakly distorted) (Fig.~\ref{fig1}d). 

As explained in the Supplementary Note 1, the Fourier transform of the magnetic correlation function associated with the LC phase is located in iridates at nuclear Bragg peaks, such as $(1,1,2+4n)$. These Bragg peaks respect both the original body-centered tetragonal structure condition $H+K+L=2n$ and the $2H+L=4m$ condition due to the $(1/2,0,1/4)$ translation. 
(The wave-vector is denoted by $\mathbf{Q}=(H,K,L)$, see Method). This produces a very specific magnetic diffraction pattern, which can be probed by magnetic sensitive diffraction technique, such as neutron scattering technique. 

\begin{figure*}
\includegraphics[width=0.9\textwidth,clip]{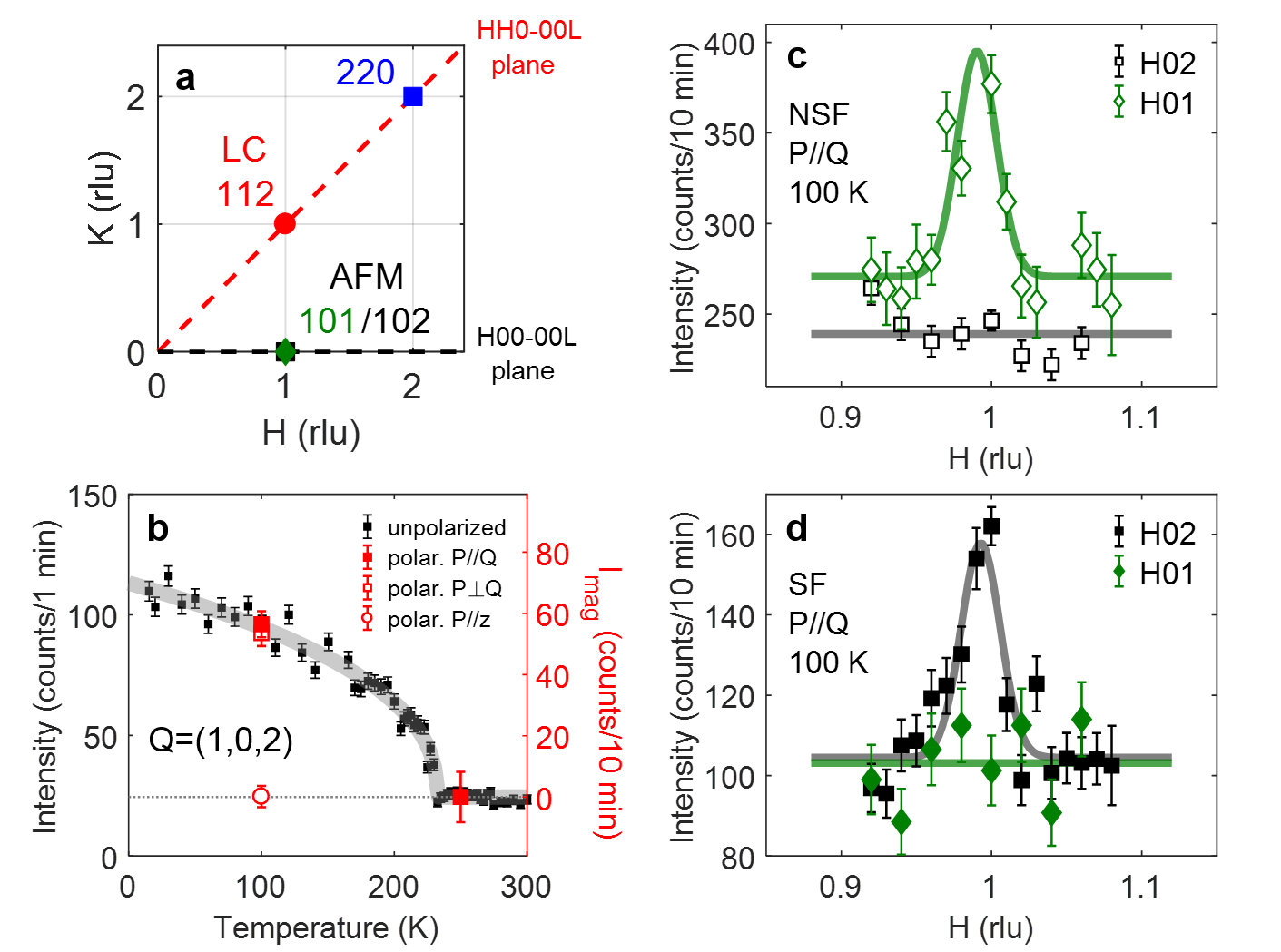}
\caption{\label{fig2}{\bf AFM order in Sr$_2$IrO$_4$}
(a) Momentum transfer $\mathbf{Q}$ positions (projected in the reciprocal $HK$-plane) investigated with neutron scattering: $(1,0,L=1,2)$ corresponds to the AFM Bragg peaks whereas  $(1,1,2)$ is where a magnetic scattering is expected for the loop current (LC) phase. (b) Thermal evolution of the peak intensity at the AFM Bragg peak $(1,0,2)$ measured by unpolarized neutron (black), and the full polarization analysis at $T=100$ and 250~K (red) (see text).  (c,d) The $(H,0,L=1,2)$ scans at 100~K for both the non-spin-flip (NSF) (empty symbols) and spin-flip (SF) (full symbols) channels (diamond symbols for $L$=1 and square symbols for $L$=2). Whilst the $(1,0,2)$ peak (black squares) shows purely magnetic behaviour, the $(1,0,1)$ (green diamonds) shows a purely nuclear signal.
}
\end{figure*}

We have investigated both the conventional AFM orders and the LC magnetic order using neutron scattering diffraction experiments. Note that the momentum transfers $\mathbf{Q}$ are different for both types of phase: $(1,0,L)$ for AFM peaks and $(1,1,L)$ for the LC peaks (see Supplementary Note 1). Three major wave-vectors positions, $\mathbf{Q}$, have been examined: their projection onto the reciprocal $HK$-plane are shown in Fig.~\ref{fig2}a. We have then studied each phase in a different scattering plane: $H00$-$00L$ for conventional AFM orders and $HH0$-$00L$ to study LC order (see Method). We have investigated two iridate samples: a pure Sr$_2$IrO$_4$ and a 7\% Rh-doped one. Small single crystals have been co-aligned to get a large enough sample mass in order to perform the polarized neutron experiment. The Supplementary Note 2 describes the sample preparation and shows magnetization measurements, which characterize the samples.

We have first studied the conventional AF-II order by repeating unpolarized neutron diffraction measurements~\cite{Ye2013,Dhital2013}. We report in Fig.~\ref{fig2}b the AFM Bragg peak at $\mathbf{Q}=(1,0,2)$ in the pure system where the magnetic Bragg intensity shows a sharp AFM transition at $T_\mathrm{N}=232$~K. The fitted critical exponent $2\beta\approx0.41$ is consistent to 0.36 in the previous report~\cite{Ye2013}. Next, we have performed polarized neutron experiment to study the $\mathbf{Q}=(1,0,L)$ Bragg peaks to disentangle the magnetic and nuclear contributions of these peaks. The polarized neutron experiment setup is presented in the Supplementary Note 3. It has been already used in previous measurements in cuprates and described in Refs.~\cite{Fauque2006,Li2008,Bourges2011,Sidis2013,Mangin2015}. For a given neutron polarization  $\mathbf{P}$, the neutron intensities, $I_\mathrm{SF}$ and $I_\mathrm{NSF}$, are measured in both spin-flip (SF) and non-spin-flip (NSF) channels, respectively. Full magnetic signal appears in the SF channel when $\mathbf{P}\parallel\mathbf{Q}$~\cite{Bourges2011,Mangin2015} whereas the nuclear intensity occurs in the NSF chanel. The Fig.~\ref{fig2}c,d shows the wave-vector scans along $H$ across the magnetic peak 
$\mathbf{Q}=(1,0,2)$ for the NSF and SF channels. That proves the magnetic origin of the peak as it is only seen in the SF channel (Fig.~\ref{fig2}d). In Fig.~\ref{fig2}b, the magnetic intensity $I_\mathrm{mag}=I_\mathrm{SF}-BG$ is determined at two temperatures, 100~K and 250~K, below and above $T_\mathrm{N}$  (where $BG$ stands for the flat background of Fig.~\ref{fig2}d). A clear magnetic intensity is sizeable at 100~K, whereas no magnetic intensity is seen at 250~K.  Further,  we report at 100~K in Fig.~\ref{fig2}b  the magnetic intensity for three different neutron polarization states: the polarizations $\mathbf{P}\parallel\mathbf{Q}$ and $\mathbf{P}\perp\mathbf{Q}$ are in the scattering plane (see Method), whereas $\mathbf{P}\parallel\mathbf{z}$ is perpendicular to the scattering plane. In the given geometry,  zero magnetic intensity with $\mathbf{P}\parallel \mathbf{z}$ proves that the AFM moments are confined in the $ab$-plane. This is in agreement with the previous studies of the neutron structure factors~\cite{Ye2013,Dhital2013}.

In the same scattering plane, the Bragg peaks at $\mathbf{Q}=(1,0,L)$ for $L=1$ and $3$ were also measured using polarized neutron. They correspond to the forbidden peaks of the original $I4_1/acd$ structure giving rise to the  space group $I4_1/a$  where the glide planes are lost~\cite{Torchinsky2015}. As it has been already discussed~\cite{Ye2013,Dhital2013}, their temperature dependence exhibits no anomaly neither at the N\'eel temperature nor at the onset of the odd-parity hidden order ($T_{\Omega}$)~\cite{Zhao2016}. The origin of that scattering is non-magnetic, as shown in Fig.~\ref{fig2}c,d. For the Bragg peak at $\mathbf{Q}=(1,0,1)$, no signal is sizeable in the SF channel and only a NSF signal is observed above the background. These results exclude the possibility of an AFM arrangement of $+-+-$ type, where the magnetic signal should be only at odd $L$-value. It casts serious doubts that such an AFM stacking can explain the origin of the SHG signal~\cite{DiMatteo2016}.  The same conclusion holds for the $++++$ arrangement, not observed in pure Sr$_2$IrO$_4$, in full agreement with the literature~\cite{Ye2013,Dhital2013,Boseggia2013}. To account for the occurrence of an odd-parity hidden order in SHG measurements in pure Sr$_2$IrO$_4$, we are left with two scenarii. Either, the missing AFM orders are light-induced metastable states during the SHG measurements, as suggested in Ref.~\citen{DiMatteo2016}, or another kind of magnetic order exists. Let us consider now this second scenario.

\begin{figure*}
\includegraphics[width=0.9\textwidth,clip]{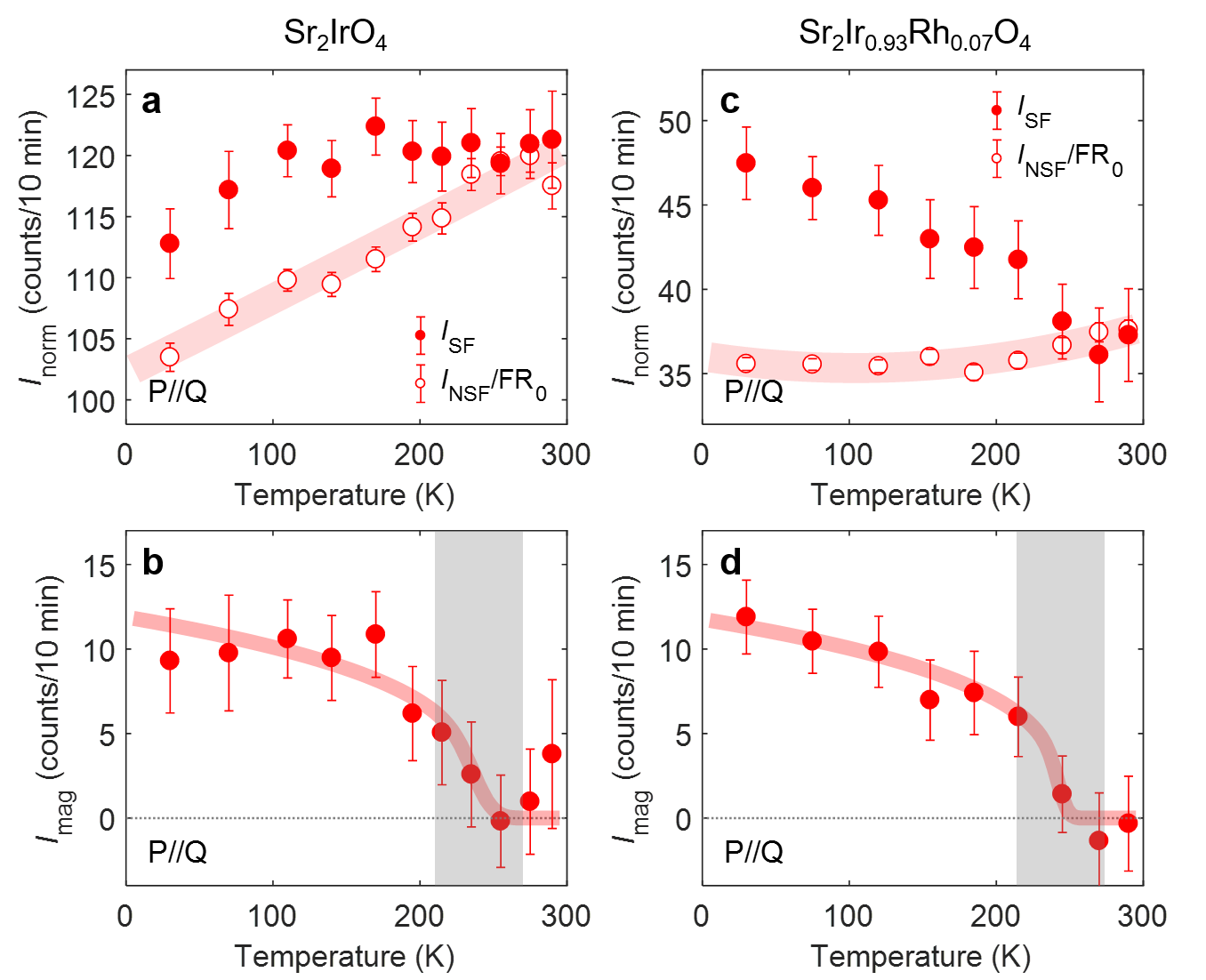}
\caption{\label{fig3}{\bf Hidden magnetic order in Sr$_2$Ir$_{1-x}$Rh$_x$O$_4$ (x=0, 0.07)}
(a,c) Temperature dependence of $I_\mathrm{SF}$ and a bare SF intensity $I_\mathrm{SF}^0 = I_\mathrm{NSF}/\mathrm{FR}_0(T)$ from the polarization leakage. The temperature-dependent bare flipping ratio $\mathrm{FR}_0(T)$ is taken from the reference Bragg peak $(2,2,0)$ and rescaled to correct the NSF intensity at $(1,1,2)$ (see Supplementary Note 4). 
(b,d) Temperature dependence of the magnetic intensity $I_\mathrm{mag}=I_\mathrm{SF}-I_\mathrm{NSF}/\mathrm{FR}_0(T)$ at the nuclear Bragg peak $(1,1,2)$. Red curves are guide to the eye and grey areas represent temperature uncertainty in the determination of $T_\mathrm{mag}$. }
\end{figure*}

The overall experimental procedure to extract the magnetic signal associated with the LC order is detailed in the Supplementary Note 4 and it follows methods established in previous works on superconducting cuprates~\cite{Fauque2006,Bourges2011}. The measured spin-flip intensity, $I_\mathrm{SF}$, is reported as a function of temperature in Fig.~\ref{fig3}a for the pure sample. It is compared with the background baseline $I_\mathrm{SF}^0$ determined from the measured NSF intensity (see the Fig.~\ref{fig3}a caption and Supplementary Note 4).  While $I_\mathrm{SF}^0$ decreases monotonically as temperature decreases, the SF intensity instead departs from $I_\mathrm{SF}^0$ below $T_\mathrm{mag}\approx240\pm30$~K. The difference, $I_\mathrm{SF}-I_\mathrm{SF}^0$, evidences a spontaneous magnetic order whose magnetic intensity, $I_\mathrm{mag}$, is reported in Fig.~\ref{fig3}b. It has a different symmetry compared to the conventional AFM phases discussed above and the position corresponds to where magnetic intensity is expected for the LC phase. For the 7\% Rh-substituted sample, the same analysis is given in Fig.~\ref{fig3}c and d. While the AFM transition is suppressed down to $T_\mathrm{N}\approx100$~K, the novel magnetic order is observed at much higher temperature $T_\mathrm{mag}\approx240\pm30$~K. Within error bars, the transition temperature, $T_\mathrm{mag}$ and its normalized magnetic intensity do not show a significant difference  between the pure system and 7\% Rh-substitution. This is consistent with the estimate of onset of the hidden order, $T_\mathrm{\Omega}$, from SHG (orange squares in Fig.~\ref{fig1}a)~\cite{Zhao2016}, which does not change appreciably with Rh substitution. Using the calibration of nuclear Bragg peaks intensities, one can deduce the magnetic cross section of the hidden magnetic order as $\sim2$~mbarns/f.u., which is less than $\sim10^{-3}$ of the strongest nuclear Bragg peak. The normalized magnitude of $I_\mathrm{mag}$ for the hidden magnetic order is similar in both samples and is $\sim5$ times smaller than the one for the AFM order, as shown in Fig.~\ref{fig3}.

{\bf Discussion}

To understand the spontaneous magnetic order at  $T_\mathrm{mag}$, a model is needed to explain the broken symmetries of the hidden order. 
Concomitant with the SHG data, time-reversal, parity inversion and 4-fold rotational symmetries are broken. It preserves the translational symmetry of the underlying lattice, as the magnetic scattering appears on top of the nuclear Bragg peak. The loop-current model~\cite{Varma1997,Simon2002} is a good candidate, since it can explain all these broken symmetries. 
Nevertheless, more detailed and quantitative study is required to establish the exact order parameter in the hidden ordered phase. The weak magnetic cross section shows how difficult it is to detect and why it was not reported with typical unpolarized neutron diffraction. Due to the experimental limitations, a precise determination of the direction of induced magnetic moments and magnetic structure requires further works.

Actually, other magnetic orders, which could potentially account for the SHG data, are not consistent with our finding as they would give rise to distinct scattering patterns. For instance, the proposed structure with $m1'$ point group~\cite{Zhao2016} yields a different relation $H+K+L=2n+1$, as the configuration is opposite when one considers the $(1/2,1/2,1/2)$ translation. This breaks the original body-centered nuclear structure \cite{Huang1994}, and would give rise to magnetic superstructure peaks at $(1,0,L=2n)$ and no magnetic intensity at any $(1,1,L)$ position. With Rh doping, one also does not observe magnetic superstructures at $(1,0,L)$ with even $L$~\cite{Clancy2014,Ye2015}. Both points dismiss the proposal of the $m1'$ point group~\cite{Zhao2016}. 

Alternatively, the SHG measurements have been re-interpreted considering conventional AFM orders~\cite{DiMatteo2016} with a different stacking along the $c$-axis.  We detail in the Supplementary Note 5 why the different AFM structures discussed in the literature cannot actually explain our observation. First, since the hidden order occurs at different $\mathbf{Q}$ positions compared to the AFM order, it is clearly distinct with the AF-I order. Second, the AFM state of $2'/m$ symmetry with a $+-+-$ stacking along the $c$-axis of AFM planes is argued to explain SHG data~\cite{DiMatteo2016}. This phase will give rise to magnetic contributions at positions such as $(1,0,L=2n+1)$ and nothing at $(1,1,L)$, where we observe the hidden magnetic intensity. Further, the non-magnetoelectric $++++$ phase (AF-II) (described above) could also account for the SHG data~\cite{DiMatteo2016}. This AF-II phase is reported in the Rh-doped system~\cite{Clancy2014,Ye2015} but is absent in the pure system~\cite{Ye2013,Dhital2013}. It would also exhibit the largest magnetic contribution at $(1,0,L=2n+1)$ as well as tiny magnetic contributions at Bragg peaks, such as $(1,1,L=2+4n)$, due to its weak ferromagnetism. This interpretation can be excluded in both samples we have studied. First, in the pure sample, the ferromagnetic order is absent (see Fig.~\ref{fig1})~\cite{Ye2013,Dhital2013} and can be only induced by an applied magnetic field of about 0.2 T (2000~Oe)~\cite{BJ2009}. Second, under Rh substitution, such a ferromagnetic order indeed develops, but only below $T_\mathrm{N}$~\cite{Clancy2014}, as it results from the canting of the AFM order, clearly lower than $T_\mathrm{mag}$. Moreover, the  neutron intensity due to ferromagnetism would be $\sim\mathrm{tan}^2\theta \leq 10^{-2}$ smaller than the AFM one, i.e at least one order of magnitude smaller than the observed magnetic scattering we report here. All these arguments allow us to rule out the weak ferromagnetism derived from the canted AFM order as a candidate to account for the observed magnetic scattering at $\mathbf{Q}=(1,1,2)$. Therefore, our observation of a magnetic signal at $\mathbf{Q}=(1,1,2)$ is not consistent with any kind of stackings along the $c$-axis of the pseudo-spin AFM orders considered to explain the SHG signal in Ref.~\cite{DiMatteo2016}. 

Using polarized neutron diffraction, we have experimentally addressed all these alternative phases and found out evidence for a hidden magnetic order in Sr$_2$Ir$_{1-x}$Rh$_x$O$_4$. It is a translation-invariant but time-reversal symmetry broken phase that is consistent with the LC order of $2'/m$ point group symmetry, concomitantly compatible with the SHG signal. In that model, all IrO$_6$ octahedra, which are the building blocks of the material, are identically decorated by the same set of staggered magnetic moments, whose magnetism cancels out on each octahedron (as depicted in Fig.~\ref{fig1}b). This magnetic order is then clearly distinct from the AFM one, where each octahedron has a single pseudo-spin on the Ir site and where the nearest octahedra have staggered moments.

In conclusion, we report the first evidence of an unconventional magnetic order in Sr$_2$(Ir,Rh)O$_4$, which breaks time-reversal symmetry but preserves translational symmetry of the underlying lattice. By analogy with superconducting cuprates, where a similar kind of order is observed, one can refer to it as an intra-unit-cell order. The new magnetic phase overlaps with parity inversion and rotational symmetry broken phase recently reported using SHG~\cite{Zhao2016}. Both observations can be described by the loop-current order~\cite{Varma1997,Simon2002}
proposed for the pseudogap state in cuprates, where it is well supported by polarized neutron measurements~\cite{Fauque2006,Bourges2011}. Further, the neutron observation in cuprates is confirmed as well by recent SHG measurements that show a global broken inversion symmetry in YBa$_2$Cu$_3$O$_{6+x}$~\cite{Hsieh2016}. This may provide more analogy between the iridates and the high-$T_\mathrm{c}$ cuprates, in spite of a different nature of $5d$ and $3d$ orbitals. A noticeable difference is here that the loop order occurs in the insulating compounds at half-filling, whereas in cuprates it is observed in the doped metallic (superconducting) state. Our report generalizes the existence of loop-current electronic states in oxides. 

{\bf Methods: Polarized neutron diffraction} 

The polarized neutron experiments were performed on the triple axis spectrometer 4F1 (Orph\'ee, Saclay, France) (see Supplementary Note 3). All measurements were done in two different scattering planes, either $H00$-$00L$ or $HH0$-$00L$, where the scattering wave-vector is quoted as $\mathbf{Q}=H\mathbf{a}^*+K\mathbf{b}^*+L\mathbf{c}^*\equiv(H,K,L)$ with $a^*=b^*=1.15~$\AA$^{-1}$ and $c^*=0.24$~\AA$^{-1}$. 
As emphasized in Supplementary Note 1, in Sr$_2$(Ir,Rh)O$_4$, the conventional AFM order is expected in the $H00$-$00L$ plane and the LC phase in the $HH0$-$00L$ plane. For each wave-vector $\bf{Q}$, the scattered intensity is measured in both spin-flip (SF) and non-spin-flip (NSF) channels. The neutron measurements have been performed with a neutron polarization $\mathbf{P}\parallel\mathbf{Q}$ where the magnetic signal appears entirely in the SF channel~\cite{Fauque2006,Bourges2011}. The AF-I order was further studied with a polarization $\mathbf{P}\perp\mathbf{Q}$ (but still within the $H00$-$00L$ plane) as well as with a polarization $\mathbf{P}\parallel\mathbf{z}$ which is perpendicular to the scattering plane (along $0K0$ in the given case). The Sr$_2$(Ir,Rh)O$_4$ single crystals preparation and caracterization are described in the Supplementary Note 2. 

\begin{addendum}
 \item We wish to thank  S. Di Matteo, D. Hsieh, B.J. Kim, M.R. Norman and C.M. Varma for fruitful discussions. We also acknowledge financial supports from the projects NirvAna (contract ANR-14-OHRI-0010) and SOCRATE (ANR-15-CE30-0009-01) of the ANR French agency. 

 \item[Competing Interests] The authors declare that they have no
competing financial interests.
 \item[Correspondence] Correspondence and requests for materials
should be addressed to Jaehong Jeong~(email: jaehong.jeong@cea.fr)
 or Philippe Bourges~(email: philippe.bourges@cea.fr).
\end{addendum}

\clearpage

\section*{\label{reciprocal_space} Supplementary Note 1}
{\bf Reciprocal space comparison with cuprates}

\begin{figure*}[h]
\includegraphics[width=\textwidth,clip]{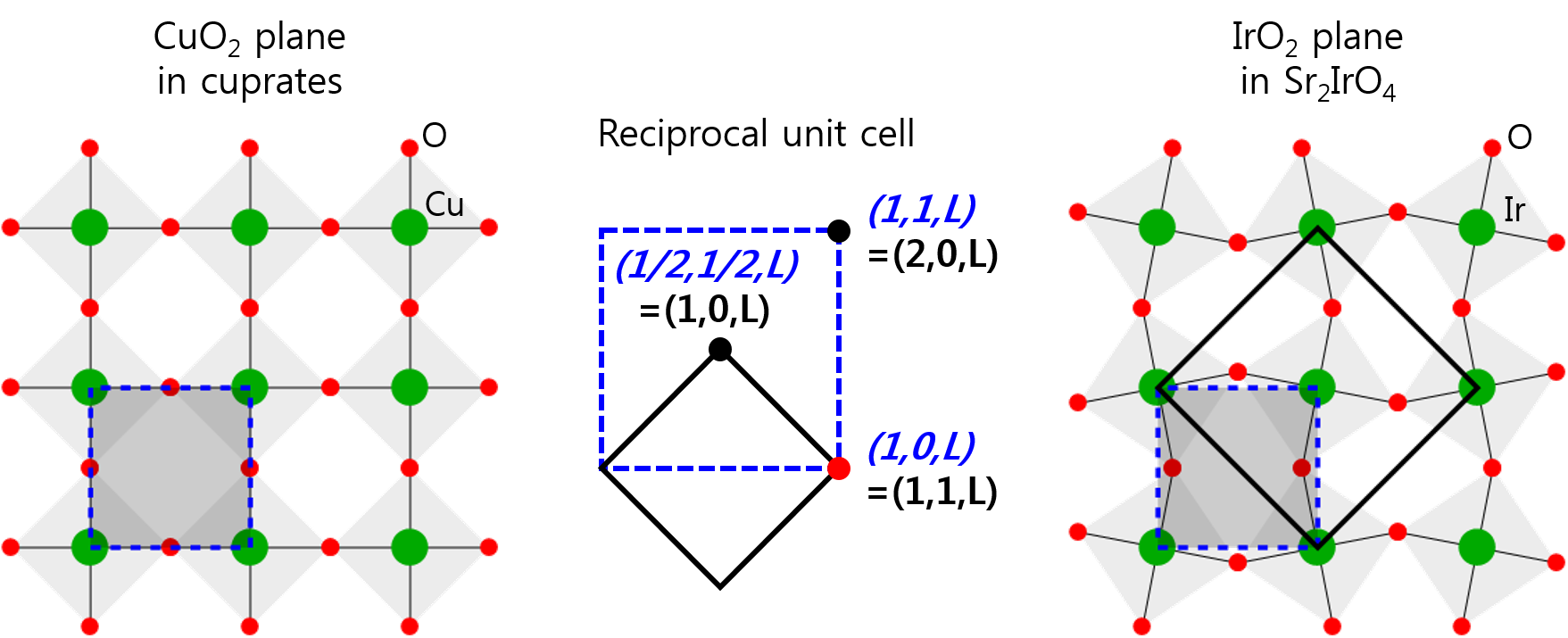}
\caption{\label{figS2}{\bf {\it M}O$_2$ plane in cuprates ({\it M}=Cu) and in Sr$_2$IrO$_4$ ({\it M}=Ir).} 
(Left figure) In cuprates, the CuO$_2$ plane is typically a weakly distorted square lattice that one generally represents as a regular square lattice, so it could be described by the minimal unit cell (dashed blue). (Right figure) In contrast, in iridates, due to a large in-plane rotation of the IrO$_6$ octahedra, the corresponding unit cell is doubled and rotated by 45$^{\circ}$ (black). (Center figure) Planar reciprocal space where a few interesting $\mathbf{Q}$-positions are compared in both systems (written in italic blue for cuprates and in black for iridates).}
\end{figure*}

The Sr$_2$IrO$_4$ exhibits a very similar in-plane structure compared to superconducting cuprates. In cuprates, corner-shared CuO$_6$ octahedra or CuO$_5$ pyramids make a (nearly) square lattice of CuO$_2$ plaquettes on the basal $ab$-plane. On the other hand in iridates, the IrO$_6$ octahedra are also corner-shared but rotated by $\theta\sim11^{\circ}$ in the basal plane. Due to this in-plane rotation, the unit cell for iridates is doubled and rotated by 45$^{\circ}$ as shown in Fig.~\ref{figS2}. Despite a difference between the spin $S=1/2$ of Cu$^{2+}$ and pseudo-spin $J=1/2$ of Ir$^{4+}$ states, the antiferromagnetic (AFM) structure is also almost the same in both systems, so the magnetic Bragg conditions are also similar. However, under the 45$^{\circ}$ unit-cell transformation, actual $(H,K,L)$ Miller indices are different. In the center of Fig.~\ref{figS2}, we compare a few interesting Bragg $\mathbf{Q}$-positions in the planar reciprocal $HK$-plane for iridates (black) and cuprates (blue). For instance, the $(1/2,1/2,L)$ for the AFM order in cuprates is transformed to $(1,0,L)$ or $(0,1,L)$ for iridates, and the $(1,0,L)$ in cuprates, where the loop-current (LC) order has been reported~\cite{Fauque2006b,Bourges2011b}, corresponds to $(1,1,L)$ for iridates.

\section*{\label{sample} Supplementary Note 2}
{\bf Coaligned single crystals and magnetization measurements}

\begin{figure*}
\includegraphics[width=0.8\textwidth,clip]{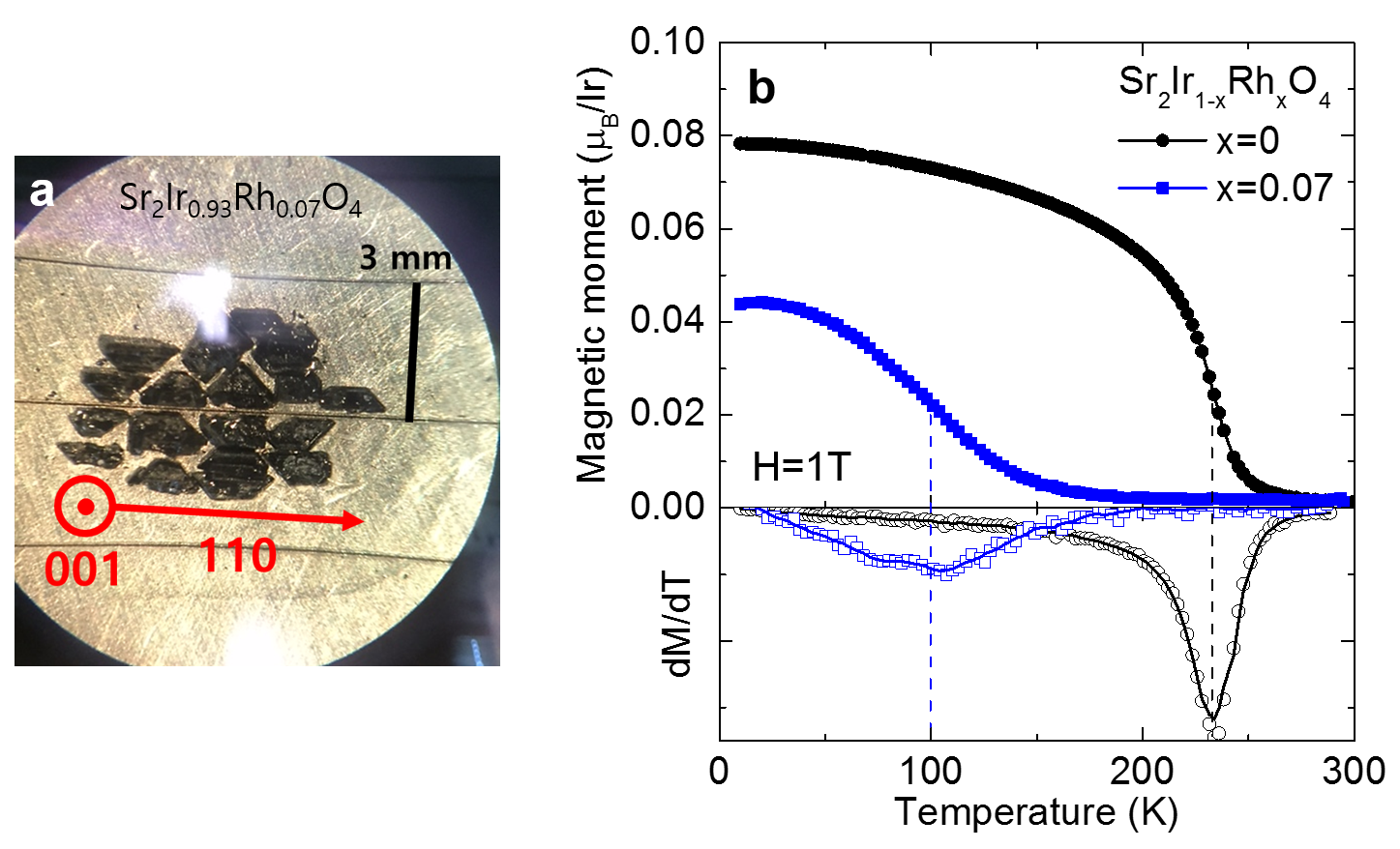}
\caption{\label{figS1}{\bf Coaligned single crystals and measured magnetic moments} 
(a) Total 18 single crystals of the 7\% Rh-doped sample were coaligned on a thin Al plate. Two major crystallographic directions are denoted: $(1,1,0)$ and $(0,0,1)$. (b) Ferromagnetic moments deduced from magnetization measurements at $H=1$~T. By substituting Rh for Ir, the magnetic transition temperature and the saturated moment are rapidly suppressed.}
\end{figure*}

We have investigated the pure and 7\% Rh-doped single crystals grown by flux method at Laboratoire de Physique des Solides (Orsay). Owing to a rectangular cuboid shape of the crystals, several tiny crystals could be coaligned in order to increase the total mass, as shown in Fig.~\ref{figS1}a. To address the conventional antiferromagnetic (AFM) order, the magnetization was measured under a magnetic field $H=1$~T using Magnetic Property Measurement System (MPMS). The pure system originally exhibits an AFM order, but it can easily show ferromagnetism,  by applying a small external magnetic field $H\approx 0.2$~T ($\simeq$2000~Oe)~\cite{BJ2009b}.  For instance, the reported magnetic moment deduced from magnetization measurements shown in Fig.~\ref{figS1}b corresponds to a net ferromagnetic moment under 1~T. This ferromagnetism originates from a canting of the  AFM structure.  The transition temperature is taken at the maximum slope in $M(T)$ curve, i.e., the minimum of the 1st derivative of magnetization, $dM(T)/dT$ shown in Fig.~\ref{figS1}b. The pure system shows a sharp transition at $T_\mathrm{N}\approx232$~K and the saturated ferromagnetic moment per Ir ion is $\sim0.08\mu_\mathrm{B}$ at 1T. On the other hand, the doped one shows a broad transition near $T_\mathrm{N}\approx100$~K and the moment is also reduced to $\sim0.04\mu_\mathrm{B}$. To map out the phase diagram (Fig.1a in the main text), the transition temperatures for different doping levels were determined from the previous literatures~\cite{Qi2012b,Ye2015b}, following the same procedure.

\section*{\label{setup} Supplementary Note 3}
{\bf Experimental neutron scattering setup}

Polarized neutron diffraction experiments were performed on the triple-axis spectrometer 4F1 located at the Orph\'ee reactor in Saclay 
(France). The polarized neutron setup was similar to the one used previously for studying cuprates~\cite{Fauque2006b,Mook2008b,Li2008b}. 
A polarized incident neutron beam with $E_i=13.7$~meV ($k_i=2.57$~\AA$^{-1}$) is obtained by a polarizing supermirror (bender) and scattered neutrons are measured with a Heusler analyzer which determines as well the final neutron polarization. A pyrolytic graphite filter is put before the bender to remove high harmonics. A small magnetic field of typically 10~G is applied using a Helmholtz-like coil. It is used to change adiabatically the direction of the neutron polarization at the sample position. A Mezei flipper is located before the sample position to flip the neutron spin. For each Bragg position, the scattered neutron intensity is measured in both spin-flip (SF) and non-spin-flip (NSF) channels, that corresponds to two different states of the Mezei flipper, flipper-off and flipper-on, respectively.  One defines the flipping ratio $\mathrm{FR}=I_\mathrm{NSF}/I_\mathrm{SF}$ of the Bragg peaks intensities in both NSF and SF channels. It determines the polarization efficiency of the apparatus. Due to unavoidable neutron polarization  leakage from the NSF to the SF channel (imperfect polarization), a value for FR was obtained between 30--50 for the pure sample and 50--65 for the doped sample. In order to measure a small magnetic signal on top of the large nuclear peak, it is essential to keep very stable and homogeneous neutron polarization through the whole measurement. Thus all the data have been measured at a fixed configuration of the spectrometer while changing the temperature.
 
\section*{\label{flipping ratio} Supplementary Note 4}
{\bf Analysis for the flipping ratio}

According to the (co-planar) loop current (LC) model, the magnetic scattering intensity should be observed at Bragg reflections $(1,1,2+4n)$, on top of the nuclear scattering.  As the LC phase respects the translation symmetry of the lattice, it corresponds to an intra-unit-cell magnetic order.  Thus, the detection of such a magnetic order critically depends on the ability to disentangle nuclear and magnetic scatterings. This difficulty can be overcomed by using  polarized neutron scattering technique~\cite{Fauque2006b,Bourges2011b}. Once the neutron spin polarization $\mathbf{P}$ is set parallel to the transferred momentum, $\mathbf{P}\parallel\mathbf{Q}$, the magnetic scattering  purely appears in spin flip (SF) channel and the nuclear one in the non-spin-flip (NSF) channel. In principle, probing separately the SF and NSF scattering channels allows one to determine the magnetic and nuclear scatterings. However, polarized neutron experiments are always limited by the quality of the neutron spin polarization, which is given by $\mathrm{FR}$, that should go to infinity for a perfectly polarized neutron beam. In practice, $\mathrm{FR}$ is finite and a fraction $\mathrm{1/FR}$ of the nuclear ({\it i.e.} NSF) scattering goes into the SF channel: it is referred to as the polarization leakage that determines a corresponding bare flipping ratio $\mathrm{FR}_0$. Another important point is to determine possible temperature dependence of the bare flipping ratio $\mathrm{FR}_0(T)$. Indeed, when changing the temperature, very tiny changes of the experimental set-up may occur\cite{Bourges2011b}, producing  a continuous drift of $\mathrm{FR}_0(T)$ that needs to be calibrated.

\begin{figure*}[t]
\includegraphics[width=12cm,clip]{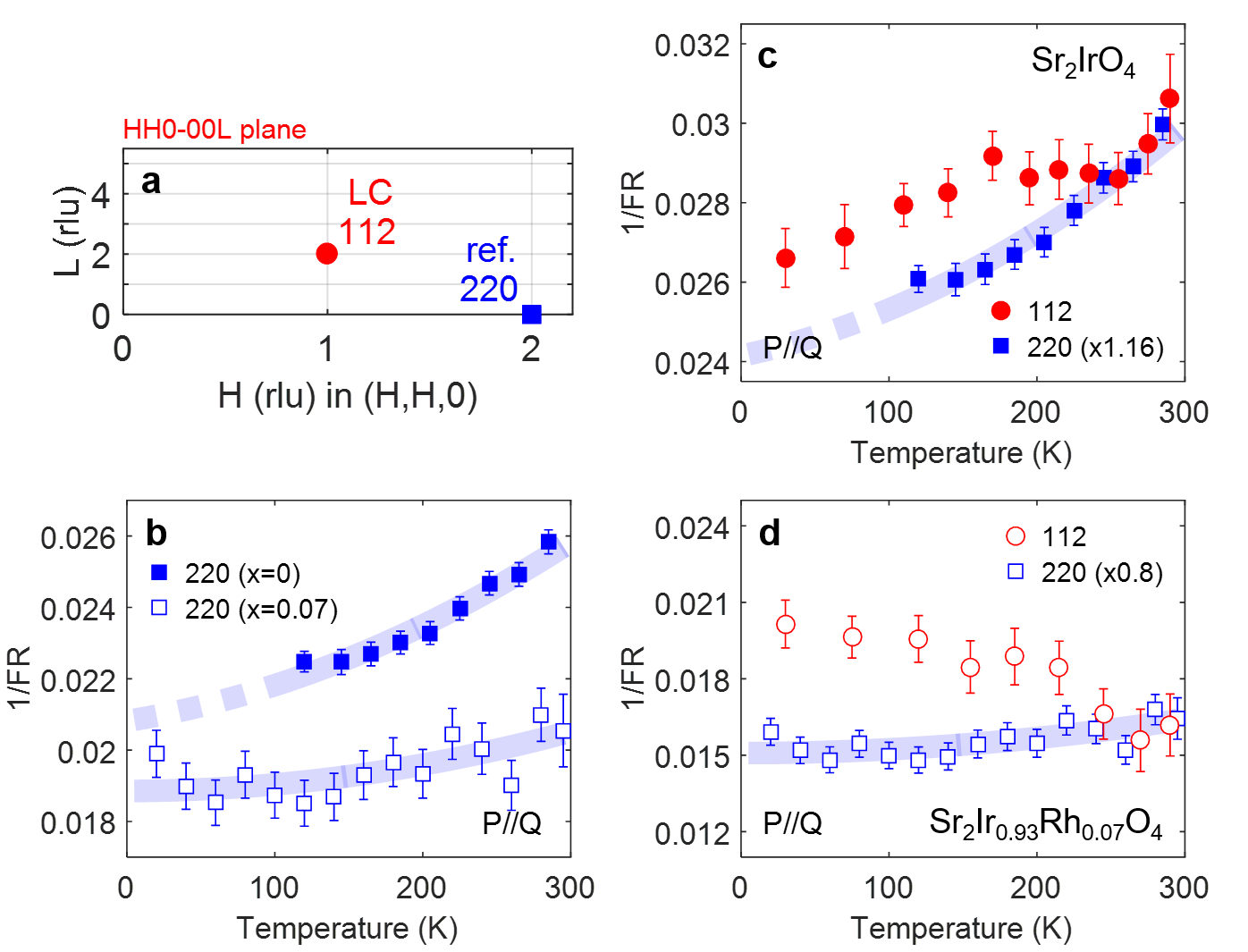}
\caption{\label{figS3}{\bf Comparison between the $(1,1,2)$ and the reference $(2,2,0)$ Bragg peaks.}
(a) (H,H,L) scattering plane where the $(1,1,2)$ (red circle) and $(2,2,0)$ (blue square) Bragg peaks positions are 
underlined. (b) 1/FR measured at the reference postion (2,2,0) for both pure (full squares) and doped (empty squares) samples, showing a small drift of 1/FR$_0$ in temperature. (c,d) The relative deviation of 1/FR at $(1,1,2)$ from the scaled $(2,2,0)$ for both systems.}
\end{figure*}

\begin{figure*}
\includegraphics[width=0.5\textwidth,clip]{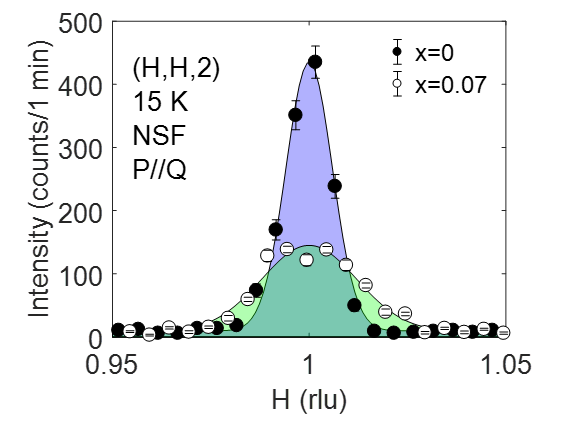}
\caption{\label{figS4}{\bf $H$-scan of the $(1,1,2)$ Bragg peak in the NSF channel.}
The integrated intensity for the NSF channel at $(1,1,2)$ in the doped sample is about 1.6 times smaller than one in the pure sample. }
\end{figure*}

In order to reveal a small magnetic signal on top of the nuclear polarization leakage in the SF channel, one has to pick up the weakest nuclear Bragg peaks (with the appropriate symmetry for the LC phase) to get the better signal-to-background ratio. In addition, the magnetic form factor is generally suppressed at a high momentum transfer. Considering these constraints, the $(1,1,2)$ reflection has been studied, which has the lowest $|\mathbf{Q}|$ of the allowed nuclear Bragg peaks $(1,1,2+4n)$ for the LC phase. Further, in order to determine the temperature dependence of the bare flipping ratio, $\mathrm{FR}_0(T)$, a non-magnetic reference Bragg peak is measured where the magnetic scattering is zero or small enough and where the spectrometer geometry is kept similar. Here, we chose the $(2,2,0)$ Bragg peak (see Fig.~\ref{figS3}a) as the reference because a magnetic signal is expected to be considerably reduced by the magnetic form factor and the spectrometer geometry is not changed much with respect to $\mathbf{Q}=(1,1,2)$. Using $\mathrm{FR}_0(T)$, obtained from the non-magnetic reference, one can then determine the intrinsic polarization leakage $I_\mathrm{SF}^0=I_\mathrm{NSF}/\mathrm{FR}_0(T)$ and next extract from the SF intensity $I_\mathrm{SF}$ the true magnetic one at $\mathbf{Q}=(1,1,2)$, $I_\mathrm{mag}=I_\mathrm{SF}-I_\mathrm{SF}^0$.  

For the pure sample, the $\mathrm{FR}_0^{220}(T)$ at the $(2,2,0)$ is $\sim$39 at 300~K and it increases up to $\sim$45 at 100~K, meaning 
that $1/\mathrm{FR}_0(T)$ (reported in Fig.~\ref{figS3}b) decreases upon cooling. That smooth decrease indicates no magnetic signal because 
one should observe a sudden increase of 1/FR in case of a magnetic order. The $\mathrm{FR}_0^{220}(T)$ for the doped sample is higher ($\sim$50) (1/FR $\sim 0.019$ reported in Fig.~\ref{figS3}b)  and much stable in temperature. That proves as well no magnetic signal at the $(2,2,0)$ peak in the doped sample. The observed temperature-dependent $1/\mathrm{FR}^{220}(T)$ can therefore be taken as the bare $1/\mathrm{FR}_0(T)$ for the polarization calibration. 

In contrast, the $\mathrm{FR}^{112}(T)$ at the $(1,1,2)$ shows a clear change of slope around 200--250~K as shown in Fig.~\ref{figS3}c,d. The $\mathrm{FR}^{112}(T)$ is $\sim$33--38 and $\sim$50--63 for the pure and doped sample, respectively. This main difference of the magnitude of FR originates from different experimental conditions such as  sample mosaicity, the number of blades on the analyzer and optimization of the guide field. In order to estimate the bare $\mathrm{FR}_0^{112}(T)$ for the $(1,1,2)$, the $\mathrm{FR}_0^{220}(T)$ is scaled by a factor that gives the same value with the $\mathrm{FR}^{112}(T)$ at high temperature above 250~K. Then, as shown in Fig.~\ref{figS3}c-d, we clearly observe a departure of $\mathrm{FR}^{112}(T)$ from $\mathrm{FR}_0^{112}(T)$ at low temperature. Using this $\mathrm{FR}^{112}_0(T)$, the bare polarization leakage, $I_{SF}^0(T) = I_{NSF}(T)/FR^{112}_0(T)$ in the SF channel can be determined at $\mathbf{Q}=(1,1,2)$. By subtracting it from the measured $I_{SF}(T)$, the magnetic intensity can then be reported as, $I_{mag}(T)=I_{SF}(T)-I_{SF}^0(T) = I_{SF}(T)-I_{NSF}(T)/FR_0(T)$. We applied this method in the data analysis presented in Fig. 3 of the main manuscript. 

Note that, for a quantitative comparison of the magnetic intensities of both samples, all raw intensities have been background-subtracted, normalized to the same monitor counts and weighted by an estimated sample weight. The Fig.~\ref{figS4} depicts scans of the nuclear Bragg peak intensity at $\mathbf{Q}=(1,1,2)$ along ${\it H}$ in both samples. Due to a broader mosaicity of coaligned crystals, the peak width is broader in the doped sample. From Fig.~\ref{figS4}, one deduces the integrated intensity of the Bragg peak, which is in the doped sample $\sim$1.6 times smaller than in the pure sample. Taking that into account, one could compare the magnetic intensity of the observed hidden order around 240~K for both the pure and doped samples.  We found a similar amount of the magnetic intensity in both samples as shown in Fig. 3b-c of the manuscript.

\section*{\label{AF models} Supplementary Note 5}
{\bf Can an antiferromagnetic structure explain the hidden magnetic order at $\mathbf{Q}=(1,1,2)$ ?}

The fact that we have ascribed the magnetic intensity at $\mathbf{Q}=(1,1,2)$ to a hidden magnetic order deserves additional comments. Briefly, the hidden order at $\mathbf{Q}=(1,1,2)$ cannot be described by any stacking of the planar pseudo-spin, $J_{eff}=1/2$, antiferromagnetic (AFM) pattern. Below, we discuss different AFM models considered in the literature: 

First, in the AFM $-++-$ model (AF-I), the magnetic intensity is expected at some $\mathbf{Q}=(1,1,L)$. However, it should be zero for $L$=2 because it does not preserve the body-centered symmetry of the unit cell, so  $H+K+L$ should be odd or $L=2n+1$. As a matter of fact, a tiny magnetic intensity (about a hundred times weaker than at $\mathbf{Q}=(1,0,2)$) has been reported at $\mathbf{Q}=(1,1,1)$ in pure ${\rm Sr_2IrO_4}$ \cite{Ye2013b}. That contribution is much weaker than at $\mathbf{Q}=(1,0,L)$ because the magnetic intensity at $\mathbf{Q}=(1,1,L)$ is proportional to the non-collinear pattern of the Ir moment. It is typically proportional to the square of the planar tilt angle of the IrO$_6$ octahedra, $\sim \theta^2$ with $\theta\simeq 11^{\circ}$. 

Second, in the AFM $++++$ model (AF-II), a magnetic intensity at $\mathbf{Q}=(1,1,2)$ could exist in principle due to the ferromagnetic component. However, this interpretation can be excluded in both samples we have studied. First, in the pure sample, the ferromagnetic order is absent (see Fig.1 of the manuscript)~\cite{Ye2013b,Dhital2013b} and can be only induced by an applied magnetic field of about 2000~Oe~\cite{BJ2009b}. Second, under Rh substitution, such a ferromagnetic order indeed develops but only below $T_\mathrm{N}$~\cite{Clancy2014b} as it results from the canting of the AFM order, clearly lower than $T_\mathrm{mag}$. We therefore rule out the weak ferromagnetism derived from the canted AFM order as a candidate to account for the observed magnetic scattering at $\mathbf{Q}=(1,1,2)$.

Third, in the hypothethic $-+-+$ model considered by Di Matteo and Norman~\cite{DiMatteo2016b}, a magnetic intensity will be also present at some $(1,1,L)$ positions. The structure factor for the $-+-+$ stacking is not zero only for even $L$ as it keeps the body-centered symmetry of the unit-cell. Further, due to the glide plane, the structure factor is proportional to $\sin (\pi/2 (H+L/2))$. Actually, it gives zero intensity at $(1,1,2)$ and then is not consistent with our finding. Further, such an AFM phase, $-+-+$, would also lead to magnetic contributions at 
$(1,0,L)$ for odd $L$, which are not observed in any neutron diffraction reports~\cite{Ye2013b,Dhital2013b} in pure ${\rm Sr_2IrO_4}$ as well as shown in the Fig. 2d of the manuscript.

Finally, one should stress that magnetic contribution from any of these AFM phases at $(1,1,L)$ would be negligible as it is systematially proportional to $\sim\mathrm{tan}^2\theta$ with $\theta\simeq 11^{\circ}$. Then, any effect around that $\bf{Q}$-position would be quantitatively too weak to be measured in our experiment. In absence of a planar tilt, it would even be strictly zero, although the signal for the LC phase will be still not zero.

\end{document}